\documentstyle[12pt,epsfig,a4]{article}

\newcommand{\be}{\begin{equation}}
\newcommand{\ee}{\end{equation}}
\newcommand{\bea}{\begin{eqnarray}}
\newcommand{\eea}{\end{eqnarray}}
\newcommand{\beast}{\begin{eqnarray*}}
\newcommand{\eeast}{\end{eqnarray*}}
\newcommand{\calm}{{\cal M}}
\newcommand{\calj}{{\cal J}}
\newcommand{\calu}{{\cal U}}
\newcommand{\calv}{{\cal V}}
\newcommand{\seff}{S\sub{eff}}

\newcommand{\bfp}{{\bf p}}
\newcommand{\bfq}{{\bf q}}
\newcommand{\dtp}{\frac{d^3p}{(2\pi)^3}}
\newcommand{\bfphi}{{\bf \Phi}}
\newcommand{\phidagphi}{\bfphi^\dagger\bfphi}
\newcommand{\tr}{\mbox{ tr }}
\newcommand{\Tr}{\mbox{ Tr }}
\newcommand{\intd}[1]{\int \! {\rm d} #1 \,}
\newcommand{\sub}[1] {_{\rm #1}}
\begin{document}
\begin{titlepage}
\begin{flushright}
DO-TH-95/08\\
May 1995
\end{flushright}

\vspace{20mm}
\begin{center}
{\Large \bf
Fermionic fluctuation corrections to bubble nucleation}

\vspace{10mm}

{\large  J. Baacke\footnote{
e-mail:~baacke@het.physik.uni-dortmund.de} and A. S\"urig
\footnote{e-mail:~suerig@het.physik.uni-dormund.de}} \\
\vspace{15mm}

{\large Institut f\"ur Physik, Universit\"at Dortmund} \\
{\large D - 44221 Dortmund , Germany}
\vspace{25mm}

\bf{Abstract}
\end{center}
We determine the fermionic corrections to the nucleation
rate of bubbles at the electroweak phase transition.
The fermion determinant is evaluated exactly and by using
the gradient expansion. The gradient expansion is found to
be a reliable approximation and is used to extrapolate to
the large values of $\nu_n = (2n+1)\pi T$ needed in the
Matsubara sum.
The contribution to effective action
is found to be negative and to be given, essentially, by the
gradient terms, the finite part of the wave function
renormalization. Only the
top quark contribution is evaluated, it is of the same order
as the Higgs- and W-boson contributions found previously, but
of opposite sign.
\end{titlepage}

%*******************************************************Introduction

\section{Introduction}
\setcounter{equation}{0}
The physics of the electroweak phase transition has been
discussed recently in various aspects \cite{Sin}.
Many subjects, as e.g. the question of
baryogenesis \cite{KuRuSha,CoKaNe}
are still controversial \cite{FaSha,GHOP}. Even the nature of the
phase transition is not known at present.
The temperature dependence of the effective potential has
been studied in perturbation theory
\cite{BFHW,FKRSh1} as well as in lattice simulations
\cite{KRSh,FHJJMC}.
If the mass of the Higgs is not too high (less than $M_W$) the phase
transition is supposed to be first order \cite{FKRSh2}.
In this case the transition
from the symmetric vacuum with massless particles to the broken
symmetry phase would proceed via bubble nucleation.
This phenomenon as well
as its cosmological aspects, has been studied by various groups
\cite{TuWeWi,LiLeTu,DLHLL}.

Part of the basic information needed in developing the
bubble formation and expansion scenario is the determination of
their nucleation rate. In the small temperature
span of $1$ GeV in which
the phase transition takes place bubbles
of various sizes can be formed.
Their nucleation rate varies over several orders of magnitude.
The basic rate is determined \cite{La,Co,CaCo} by the classical
minimal bubble
action. Its exponential is the tunneling rate. The semiclassical
reaction rate includes, however, also preexponential factors,
the fluctuation determinants, determined by the fluctuation
of W boson, Higgs and fermion fields in the background of the
minimal bubble profile. The bosonic
fluctuations have been computed recently
\cite{KriLaSch,Baa2} and found to yield sizeable suppression
factors, the one-loop effective action (or equivalently the free energy
divided by the temperature) being of the
same order as the classical bubble action.  In
the high temperature theory,
obtained by retaining only the Matsubara frequency $0$, fermions do not
contribute. However, recent determinations of the fermionic
contribution to the sphaleron rate \cite{DyaGoe} let us
expect that at least the top quark will influence the
transition rate in an essential way. Of course we have to perform
such a computation in the four-dimensional finite temperature
quantum field theory, i.e. by summing over all Matsubara
frequencies.

The plan of this paper is as follows:
in the next section we will introduce the model and set up
the basic relations for the bubble nucleation rate.
The fermion determinant is defined in section 3.
In section 4 we will
discuss the leading terms, of first and second order
in the external field vertex function.
Part of their contribution is contained
already in the effective potential and should not be included again.
This renders the discussion in that section very lengthy and pedantic.
In section 5 we present the computation of the finite higher order
contributions to fermionic effective action. The results
of an exact numerical computation are compared with an analytic
approximation based on the gradient expansion.
The latter one is used, then, to obtain the actual results which are
presented and discussed in section 6.

\section{Basic relations}
\setcounter{equation}{0}

As a basis for the computation of fluctuations we need an action
which describes the nucleation of minimal bubbles and which can be
used to compute the bubble profiles.
This must necessarily be a finite temperature action which
describes the essential features of the phase transition.
We will be using here an action obtained in the
electroweak theory by evaluating the
one-loop effective potential \`a  la Coleman-Weinberg \cite{CoWe}.
This has been computed by various authors
\cite{KiLi,Sha,AnHa}; here we use the
formulation of Dine et al. \cite{DLHLL}.
The finite temperature action was found by these authors
- for the temperatures relevant to the phase transition - to be
well represented by its high temperature limit. Though we will
compute the fermion determinant at finite temperature, it should be
a good approximation to compute the bubble profile from this
approximate action the mores so as the one-loop corrections
will be computed exactly and their contribution to
the approximate one-loop
effective potential will be subtracted at the end in order to avoid
double-counting (see below). This way of determining the
`classical' profile from an action
including already one-loop corrections
has been discussed critically by Weinberg \cite{Wei}; we think that
this intermediate approach nevertheless yields relevant information,
namely the order of magnitude and sign of the corrections as well
as their structure and relative importance.
\bigskip

The basic Euclidean finite temperature action then given by
\footnote{
We leave out the gauge fields here, since they
do not appear in the bubble background field configuration
and since we are
not going to compute their fluctuation determinant.}
\be \label{htac}
S_{\rm ft}  = \int_0^\beta d\tau
\int d^3x \left[
\frac{1}{2}(\partial_\mu\Phi)^\dagger (\partial_\mu \Phi)
 + V_{ht}(\phidagphi)\right] + S_F \; .
\ee
$\bfphi$ is the complex doublet of Higgs fields. Here this field
will always occur as a background field describing the
minimal bubble. It can be parametrized then as
\be \label{Higgs1}
\bfphi (\vec x)= v_0 \Phi (\vec x)
\left (\begin{array}{c} 0 \\ 1 \end{array} \right ) \; .
\ee
$V_{ht}$ is the high temperature potential which includes the
one-loop effective potential in the high temperature
approximation
\be \label{htpot1}
V_{\rm ht}(\phidagphi) =  D(T^2-T_0^2)\phidagphi-ET
\phidagphi^{3/2}+\frac{\lambda_T}{4}\phidagphi^2
\; .
\ee Its parameters are given - for $\Theta_w=0$ -
by
\bea \label{coeffs}
D&=& (3m_W^2+2m_t^2)/8v_0^2 \nonumber \\
E&=& 3 g^3/32 \pi \nonumber \\
B&=& 3 ( 3m_W^4 - 4 m_t^4)/64\pi^2v_0^4 \nonumber\\
T_0^2&=& (m_H^2-8 v_0^2 B)/4D  \\
\lambda_T&=& \lambda -3(3m_W^4\ln\frac{m_w^2}{a_BT^2}
-4 m_t^4 \ln\frac{m_t^2}{a_F T^2})/16 \pi^2 v_0^4
\eea
with
$\ln a_B=2\ln 4\pi-2\gamma$ and $\ln a_F=2 \ln \pi - 2\gamma$.
\bigskip

For $T > T_0$ the potential has a minimum at $|\bfphi|=0$
corresponding to the symmetric phase and a second minimum at
\be
|\bfphi|= \tilde v (T) = \frac{3 E T}{2\lambda}+
\sqrt{\left(\frac{3ET}{2\lambda}\right)^2+v^2(T)}
\ee
where
\be
v^2(t)= \frac{2D}{\lambda_T}(T_0^2-T^2) \; .
\ee
This minimum is degenerate with the one at $\bfphi =0$
at a temperature defined implicitly by
\be
 T_C = T_0/\sqrt{1 - E^2/D\lambda_{T_C}} \; .
\ee
$T_C$ marks the onset of bubble formation by
thermal barrier transition.
\bigskip

The fermion action $S_F$ can be written - for vanishing
gauge fields and for Higgs field configurations of the form
(\ref{Higgs1}) - in terms of four-component Dirac spinors as
\be
S_F = \int_0^\beta d\tau \int d^3 x \left [
\sum_{f}(\overline \psi^f  \gamma_\mu \partial_\mu \psi^f)
- \sum_{ff'}
g_Y^{ff'} v_0 \Phi \overline \psi^f \psi^{f'} \right ]
\ee
where the Yukawa couplings are related to the
fermion mass matrix via $g_Y^{ff'} v_0 = m^{ff'}$. The sum over
$f$ is over flavors and colors. Here we will consider
only the contribution of the top quark. This has been done already
in the high temperature action given above for the reason that its
contribution is much larger than the one of lighter
quark and lepton fields. This will also be the case for the
exact one-loop action.
\\ \bigskip

The process of bubble nucleation is - within the approach
of Langer \cite{La} and Coleman and Callan
\cite{Co,CaCo}, followed by the work of
Affleck \cite{Af}, Linde \cite{Li} and others
 - described by the rate
\be \label{rate}
\Gamma/V = \frac{\omega_-}{2 \pi} \left (
\frac{\tilde S}{2\pi}\right )^{3/2}\exp(-\tilde S)~
\left(\frac{{\cal J}_F}{{\cal J}_B}\right)^{1/2}
\; . \ee
Here $\tilde S$ is the high-temperature action, Eq. (\ref{htac}),
with the new rescaling, minimized by a classical
$\tau$ independent minimal
bubble configuration (see below). ${\cal J}_{F/B}$ are the
fermionic and bosonic fluctuation
determinants which describe the next-to-leading part of the
semiclassical approach. ${\cal J}_F$ - whose
computation is the aim of this work - will be defined below;
its logarithm is related to the fermionic one-loop
effective action by
\be \label{sefff}
\seff^F = - \frac{1}{2} \ln {\cal J}_F \; .
\ee
Finally $\omega_-$ is the absolute value of
the unstable mode frequency.

The classical bubble configuration is described by
a vanishing gauge field and a real $\tau$ independent
 spherically symmetric Higgs field.  For the bubble
configuration we make the Ansatz
\be \label{bubconf}
\bfphi(\vec x) = \tilde v (T) \phi(r)
 \left ( \begin{array}{c}0 \\ 1 \end{array}
\right )
\ee
where we have rescaled Eq. (\ref{Higgs1}) via
 $v_0 \Phi(r) = \tilde v(T) \phi(r)$.
In all our numerical computations we will use the
scale $(g_w \tilde v (T))^{-1}$ for the coordinates.
Defining the rescaled high temperature potential as
\be
\tilde V_{ht} = \frac{\lambda_T}{4 g^2}
\left[ \phi^4 - \epsilon \phi^3 + (\frac{3}{2}\epsilon -2) \phi^2
\right ]
\ee
with
\be \label{epsdef}
\epsilon = \frac{4ET}{\lambda_T \tilde v(T)} = \frac{4}{3}
\left ( 1 - \frac{v^2(T)}{\tilde v(T)^2} \right )
\ee
and the high temperature coupling
\be
\tilde g_3^2(T) = \frac{g_w T}{\tilde v(T)}
\ee
the bubble action is  given by
\be \label{Stdef}
\tilde S = \frac{4 \pi}{\tilde g_3^2(T)} \int_0^\infty r^2 dr
\left [ \frac{1}{2} \left(\frac{d\phi}{dr} \right)^2 +
\tilde V_{ht}(\phi)\right ] \; .
\ee
It is minimized if $\phi(r)$ is a solution of
 the associated Euler-Lagrange equation
\be \label{Clbub}
-\phi''(r)-\frac{2}{r}\phi'(r)+\frac{d V_{ht}}{d\phi(r)} = 0
\ee
with the boundary conditions
\be
\lim_{r\to\infty}
\phi(r)=0 ~~
 {\rm and} ~~ \phi'(0)=0 \; .
\ee
The bubble configuration varies from small thick wall bubbles to
large thin wall bubbles in the narrow range ($\simeq 1$ GeV) between
$T_0$ and $T_C$, both of order $100$ GeV.
We will use the variable
\be \label{ypsdef}
y = 3 (1 -  \epsilon/2) ~, ~~~0 < y < 1
\ee
instead of $T$ to parametrize this range of temperatures.

%---------------------------------------------------------------------

\section{The fermionic fluctuation determinant}
\setcounter{equation}{0}

The fermionic action $S_F$ of quarks can be rewritten
in four component Dirac notation and for time-independent
background configurations as
\bea
S_F &=&  \int_0^\beta d\tau \int d^3 x
\overline \psi ( \gamma_\mu \partial_\mu - m(\vec x))\psi \\
&=& \int_0^\beta d\tau \int d^3 x \psi^\dagger
( \partial /\partial \tau - H ) \psi
\eea
where
$m(\vec x) = g_Y \tilde v (T) \phi(r)$ and
\be
H = \gamma^0( -i \vec \gamma \vec \nabla + m(\vec x)) \; .
\ee
The field fluctuations are subject to antiperiodic boundary conditions
at $\tau=0$ and $\tau=\beta$, i.e.
\be
\psi(\vec x,\beta)=-\psi(\vec x, 0)
\ee
which determines their frequencies to be
the Matsubara frequencies
$\nu_n = (2 n +1)\pi T $ with integer n, $-\infty < n < \infty$.
Integrating out the fermion field leads to the fermionic prefactor
\be
\calj_F^{1/2} = \frac{\prod_{n\alpha}(i \nu_n+\omega_\alpha)}
{\prod_{n,\alpha}(i\nu_n+\omega^0_\alpha)}
\ee
where $\omega_\alpha$ denotes the eigenvalues of $H$.
Using the fact that these eigenvalues occur in
pairs $\pm \omega_\alpha$ we can rewrite this as
\be
\calj_F = \left ( \frac{\prod_{n\alpha}(\nu_n^2+\omega_\alpha^2)}
{\prod_{n,\alpha}(\nu_n^2+(\omega_\alpha^0)^2)}
\right ) \; .
\ee
The fermion contribution to the
effective action is therefore given by
\be \label{effac}
\seff^F= - \frac{1}{2} \ln \calj_F
= - \frac{1}{2}\sum_{n=-\infty}^\infty
 \ln \det \left (\frac{\nu_n^2 + \calm}{\nu_n^2+\calm^0}
\right ) \; .
\ee
Here $\nu_n = (2n+1) \pi T$, the fluctuation operators
$\calm$ and $\calm^0$ are defined as
\bea
\calm &=& H^2 = -\Delta + \calv (\vec x) \nonumber \\
\calm^0& =& (H^0)^2 =-\Delta
\eea
and
\be
\calv = \left \{ \begin {array}{cc} m^2(\vec x) & -
 i\vec \sigma \vec \nabla m(\vec x)
\\ i\vec \sigma \vec \nabla m(\vec x) &
m^2(\vec x)  \end{array} \right \} \; .
\ee
A method for computing such fluctuation determinants numerically
has been described recently \cite{BaaKi}.
Before we discuss the numerical
part of the computation we have to ensure that the quantities
we are going to compute are finite. The effective action as defined
formally in Eq. (\ref{effac}) is divergent. This is easily seen
by expanding it w. r. t. the potential $\calv$.
One generates then the series of Feynman graphs depicted in Fig. 1,
of which the first and second order graph are divergent.
We will see that our numerical method allows to separate these
two graphs from the remaining series which can be computed exactly
and is finite. The divergences of the two leading graphs are
obviously those of ordinary perturbation theory. Their divergent
parts can be cancelled by the counterterms of the $T=0$ theory.
This will be decribed in the next section, following the work of
Refs. \cite{DoJa} and \cite{DLHLL}.

%--------------------------------------------------------------------

\section{Renormalization}
\setcounter{equation}{0}

We note first that the effective potential is renormalized usually
at $T=0$ in such a way that the vacuum expectation
value $v_0$ and the Higgs mass,
i.e. the second derivative at the broken symmetry minimum,
are kept at their physical values. This will determine the
counterterms in the effective action. We will, on the other hand,
expand the effective action around the symmetric vacuum
defined by $\bfphi =0$. The discussion of renormalization
- especially at finite temperature - is therefore somewhat
cumbersome;
we will follow all steps very explicitly, though this may seem
somewhat pedantic.  \\ \\
In order to cover both kinds of expansion
we rewrite the effective action  as
\be \displaystyle
\seff =-\frac{1}{2}\ln \calj
= - \frac{1}{2}  \sum_{n=-\infty}^{+\infty}
\ln \left ( \frac{\det (\nu_n^2-\Delta+\mu_F^2+\calu)}{\det (\nu_n^2
-\Delta+\mu_F^2)} \right )
\ee
where now
$\mu_F=m_F= g_Y v_0$ if one expands around the $T=0$ broken
symmetry vacuum and $\mu_F =0$ if one expands around the high
temperature symmetric phase. The $4 \times 4$ matrix $\calu$ is
then given by
\be
\calu=\calv - \mu_F^2=\left \{ \begin{array}{cc}
m_F^2 \Phi^2 -\mu_F^2 & - im_F \vec \sigma \vec \nabla \Phi \\
i m_F \vec \sigma \vec \nabla \Phi & m_F^2 \Phi^2 -\mu_F^2
\end{array} \right \} \; .
\ee
The field $\Phi$ is normalized in such a way
that it takes the value $1$
if the Higgs field is at its $T=0$ vacuum expectation value.

In the following we will discuss the contributions of first and
second order in the external potential both at $T=0$ with massive
propagators and at large $T$ with massless propagators.
Anticipating that we work at finite temperature we replace the
space-time integration by $\beta\int d^3x$. Furthermore, for the
second order diagrams we separate space and loop momentum integration
formally, though the external momentum $q$ acts as an operator in
$x$ space. This compact notation should not lead to confusion.
\bigskip

For the first order diagram at $T = 0$ we have
\be
\seff^{(1)}(\Phi,0)=
-\frac{1}{2} 4 \beta \int d^3x  (m_F^2|\Phi(x)|^2-\mu_F^2)
\int \frac{d^4p}{(2\pi)^4}\frac{1}{p^2+\mu_F^2}  \; .
\ee
Using a 4 - momentum cutoff $\Lambda$ we get
\be
\seff^{(1)}(\Phi,0)=
-\frac{1}{8\pi^2} \beta \int d^3x (m_F^2|\Phi(x)|^2-\mu_F^2)
(\Lambda^2 -\mu_F^2 \ln (\Lambda^2/\mu_F^2)) \; .
\ee
Before we dispose further by introducing appropriate
counterterms we turn to the second order contribution at
$T=0$.
We have to evaluate
\bea
\seff^{(2)}(\Phi,0)
&=&  \frac{1}{4}
4  \beta\int d^3x((m_F^2 |\Phi(x)|^2-\mu_F^2)^2+ m_F^2
|\nabla \Phi(x)|^2)  \\ \nonumber &
&\times \int \frac{d^4p}{(2\pi)^4}\frac{1}{p^2+\mu_F^2}
\frac{1}{(p+q)^2+\mu_F^2} \; .
\eea
A standard computation, using again a 4-momentum cutoff
$\Lambda$ leads to
\bea
\seff^{(2)}(\phi,0)
&=&\frac{1}{16 \pi^2} \beta\int d^3x
 ((m_F^2 |\Phi(x)|^2-\mu_F^2)^2
+m_F^2|\nabla \Phi|^2) \\ \nonumber
&&\times(\ln \left(\frac{\Lambda^2}{\mu_F^2}\right)-1
-\int_0^1d\alpha \ln (1 + \frac{q^2}{\mu_F^2}\alpha(1-\alpha))) \; .
\eea
Putting this together with the first order contribution
and requiring all corrections to the
potential and the wave function renormalization to vanish at
$q^2=0$ we find that
we need a counterterm
\bea
\seff^{\rm c.t.} &=& \frac{\Lambda^2}{8 \pi^2}
\beta \int d^3x (m_F^2(|\Phi(x)|^2-1)) \\ \nonumber
&&-\frac{1}{16 \pi^2} (\ln \left (\frac{\Lambda^2}{m_F^2} \right )-1)
\beta\int d^3x(m_F^4 (|\Phi(x)|^4 - 1) +m_F^2 |\nabla \Phi(x)|^2)
\\ \nonumber
&& -\frac{1}{8\pi^2}m_F^4\beta\int d^3x (|\Phi(x)|^2-1)
 \; .
\eea
If the first and second order diagram are evaluated in the symmetric
vacuum we find
\bea
\seff^{(1+2,S)}(\Phi,0) &=& -\frac{\Lambda^2}{8\pi^2}
\beta \int d^3 x m_F^2|\Phi(x)|^2 + \frac{1}{16 \pi^2}
(\ln \left( \frac{\Lambda^2}{q^2}
\right ) +1)\\ \nonumber
&&\times \beta \int d^3x(m_F^4 |\Phi(x)|^4 +
m_F^2|\nabla\Phi(x)|^2)\;.
\eea
Leaving out the field-independent terms - which are due to a difference
in the vacuum energy density - we get
\bea \label{totzero}
\seff^{(1+2,S)}(\Phi,0)+\seff^{\rm c.t.}&=&
-\frac{1}{16 \pi^2}(\ln \left( \frac{q^2}{\mu_F^2}
\right )-2) \beta\int d^3x
 (m_F^2 |\Phi(x)|^4 +m_F^2 |\nabla \Phi(x)|^2)
\nonumber \\
&&-\frac{1}{8\pi^2}\mu_F^2\beta \int d^3x m_F^2|\Phi(x)|^2
\eea
which is to be evaluated with the background profile. The last term
is already contained in the high temperature potential (\ref{htpot1}),
it occurs in the coefficient $-D T_0^2 =-(m_H^2-8v_0^2B)/4$
via the top contribution to $B$.
\bigskip

Turning now to the first order diagram at finite temperature
we have to evaluate
\be
\seff^{(1)}(\Phi,T)
= -  \frac{1}{2}
4  \beta \int d^3x(m_F^2|\Phi(x)|^2 - \mu_F^2)
T \sum_{n=-\infty}^{+\infty}
\int \dtp \frac{1}{\nu_n^2 + \mu_F^2+\bfp^2}\; .
\ee
In order to separate $T=0$ and finite temperature contributions
we use \cite{DoJa}
\bea
&&T \sum_{n=-\infty}^{\infty} \frac{1}{((2n+1)\pi T)^2+ \bfp^2
+\mu_F^2}\\ \nonumber
&=& \frac{1}{2E_F} -\frac{1}{E_F}\frac{1}{\exp(E_F/T)+1}
\eea
with $E_F=\sqrt{\bfp^2+\mu_F^2}$.
In the last line the second term vanishes as $T \to 0$, so the
first one represents the $T=0$ contribution which we have considered
earlier. Inserting the second part into the expression for
${\cal J}^{(1)}$ we find the finite temperature part
\be
\Delta \seff^{(1)} (\Phi,T)
= 2 \beta \int d^3x (m_F^2 |\Phi(x)|^2 - \mu_F^2)
\int \dtp \frac{1}{E_F}\frac{1}{\exp(E_F/T)+1} \;.
\ee
We have to evaluate this expression for a bubble
in the symmetric vacuum where $\Phi \neq 0 $ only  locally and
where $\mu_F=0$ at spatial infinity. Then $E_F = |\bfp|$ and we find
\be
\Delta \seff^{(1)} (\Phi,T) =
\frac{T^2}{12} m_F^2\beta\int d^3x |\Phi(x)|^2 \; .
\ee
This contribution is already taken into account in the
$T^2$ term of the three dimensional high temperature
action (\ref{htac}). Therefore the finite temperature part
of the first order tadpole diagram has to be omitted entirely.
\bigskip

As the second order contribution at finite $T$
we have to evaluate
\bea
\seff^{(2)}(\Phi,T)&=&
\frac{1}{4} \beta\int d^3x 4(m_F^4|\Phi(x)|^4+m_F^2
|\nabla \Phi(x)|^2) \\ \nonumber
&&\times T\sum_{n=-\infty}^{+\infty}
\int \dtp \frac{1}{(\nu_n^2+\bfp^2)(\nu_n^2+(\bfp+\bfq)^2)}\; .
\eea
Momentum integration and Matsubara
frequency summation can be carried out via
\bea
&&T\sum_{n=-\infty}^{+\infty}
\int \dtp \frac{1}{(\nu_n^2+\bfp^2)(\nu_n^2+(\bfp+\bfq)^2)}
 = \nonumber \\ \nonumber
&&T\sum_{n=-\infty}^{+\infty} \int_0^1 d\alpha
\int \dtp \frac{1}{(\nu_n^2+\bfp^2+ \alpha(1-\alpha) \bfq^2)^2}
= \\
&&
T\sum_{n=-\infty}^{+\infty}\int_0^1d\alpha
\int \dtp \frac{-d}{d(\bfp^2)}
\frac{1}{\nu_n^2+\bfp^2+\alpha(1-\alpha)\bfq^2}
= \\ \nonumber
&&\frac{1}{\pi^2T} \int_0^1d\alpha
\int \dtp \frac{-d}{d(\bfp^2)}\frac{\pi^2T}{2E_\alpha}
\tanh(E_\alpha/2T)
\eea
where $E_\alpha^2=\bfp^2+\alpha(1-\alpha)\bfq^2$.
The $T=0$ part may be recovered by performing the limit
$T\to 0$. Subtracting this part we find for the finite
temperature supplement
\bea
&&\int_0^1 d\alpha \frac{1}{2 \pi^2}\int_0^\infty dp p^2
\frac{-d}{2 p dp}\frac{-1}{E_\alpha}\frac{1}{\exp(E_\alpha/T)+1}
= \\ \nonumber
&&\int_0^1 d \alpha \frac{-1}{4\pi^2}
\int_0^\infty \frac{dp}{E_\alpha}\frac{1}{\exp(E_\alpha/T)+1}
\eea so that
\bea \label{secondT}
\Delta \seff^{(2)}(\Phi,T) &=&- \frac{1}{4\pi^2}
\beta \int d^3x (m_F^4|\Phi(x)|^4
+m_F^2|\nabla \Phi(x)|^2) \\ \nonumber
&&\times
\int_0^1d\alpha\int_0^\infty
\frac{dp}{E_\alpha}\frac{1}{ \exp(E_\alpha/T)+1)} \; .
\eea
Part of this term is already contained in the high temperature
effective action.
 The momentum integral has been considered by Dolan and
Jackiw \cite{DoJa};
it can be expanded to leading order in T, i.e. up to terms of
order $1/T^2$ as
\be
\int_0^\infty \frac{dp}{E_\alpha}\frac{1}{\exp(E_\alpha/T)+1}
=
-\frac{1}{4} ( \ln\left(\frac{\alpha(1-\alpha)q^2}{\pi^2T^2}
\right ) +2 \gamma) + O(q^2/T^2)
\ee
If this is inserted into the previous equation we find
\bea
\Delta \seff^{(2)}(\Phi,T)
&=&\frac{1}{16\pi^2}\int_0^1 d \alpha
( \ln\left(\frac{\alpha(1-\alpha)q^2}{\pi^2T^2} \right)
+2\gamma ) \\ \nonumber
&&\beta \int d^3x (m_F^4|\Phi(x)|^4
+m_F^2|\nabla \Phi(x)|^2) \; .
\eea
The integration over $\alpha$ can then be performed,
replacing the first parenthesis by
\be
( \ln\left(\frac{q^2}{\pi^2T^2}\right) -2
+2\gamma) \; .
\ee
If this is added to the zero temperature result
(\ref{totzero}) the term $\ln (q^2/m_F^2)-2$ in this equation
gets replaced by
$\ln (T^2/m_F^2)+2\ln \pi -2\gamma$. This term appears in the
high temperature potential in $\lambda_T$.
\bigskip

Collecting from Eqs. (\ref{totzero})
and (\ref{secondT}) the terms which have not yet been included into
the high temperature potential we find the following renormalized
contribution of the first and second order Feynman graphs:
\bea \label{DeltaS12}
\Delta S^{(1+2)}_{\rm ren}(\Phi,T)&=&
- 3 \beta \frac{m_t^4}{16 \pi^2}\int \frac{d^3q}{(2\pi)^3}
\left (|\widetilde{\Phi^2}(q)|^2 + \frac{q^2}{m_t^2}
 |\tilde\Phi(q)|^2 \right) \\
\nonumber
&&\left(\ln \left(\frac{q^2}{m_t^2}\right)-2
+4\int_0^1d\alpha\int_0^\infty\frac{dp}{E_\alpha}
\frac{1}{\exp(E_\alpha/T)+1} \right ) \\ \nonumber
&& - 3\beta \frac{m_t^4}{16\pi^2} \ln \left(\frac{m_t^2}{a_FT^2}\right)
\int \frac{d^3q}{(2\pi)^2} |\widetilde{\Phi^2}(q)|^2 \; .
\eea
Here we have taken into account the color factor $3$, we have
replaced $m_F$ by $m_t$ and we have replaced the formal
$x$ integration by a $q$ integration in order to deal with
the q dependence of the kernels correctly. Note that
$\Phi^4(x)$ becomes here the square of the Fourier transform
of $\Phi^2$.

%------------------------------------------------------------------

\section{Calculation of the finite part of the effective action}
\setcounter{equation}{0}

What remains to be evaluated is the sum of all Feynman graphs
of order $3$ and higher. This contribution is finite, we denote
it as $\seff^{\overline{(3)}}(\phi)$. We present an exact
numerical computation, using a general theorem on functional
determinants \cite{Col} and an analytic approximation based on
the gradient expansion. The results of the two methods will be compared
at the end of this section.
\subsection{Numerical computation}
As mentioned above the fermionic
one-loop effective action at finite temperature can be
written - including the color factor 3 - as
\begin{equation}
\label{effaction}
S\sub{eff} = -\frac{3}{2} \sum_{n=-\infty}^{\infty} \ln \det
\left( \frac{-{\bf \Delta} + \nu_n^2 + {\cal V}}
{-{\bf \Delta} + \nu_n^2}
\right) = - 3 \sum_{n=0}^{\infty} \ln {\cal J}(\nu_n).
\end{equation}
As the background field is spherically symmetric the determinant can
be decomposed into its partial wave contributions. This is readily
done by introducing the usual spinors for given $j$ and $l=j\pm1/2$
as given in the textbook of Bjorken and Drell \cite{BjDr}.
One finds then
\begin{equation}
S\sub{eff} = - \sum_{n=0}^{\infty} \sum_{l=0}^{\infty} 6(2l+1) \ln
{\cal J}_l(\nu_n) \; .
\end{equation}
Here the the partial wave determinants ${\cal J}_l(\nu)$
are defined as
\[
{\cal J}_l(\nu)= \det \left(
\frac{{\bf M_l} + \nu^2}{{\bf M_l^0}+ \nu^2}
\right)
\]
with the partial wave fluctuation operators
\begin{eqnarray}
{\bf M_l} &=& {\bf M_l^0} + {\bf V}(r)\nonumber \\
{\bf M_l^0} &=&
-{\bf 1}\left (\frac{d^2}{dr^2} + \frac{2}{r} \frac{d}{dr}\right)
+ \frac{1}{r^2}\left\{\begin{array}{cc}l(l+1)&0\\0&
(l+1)(l+2)\end{array}\right\}\\
{\bf V}(r)&=&\left\{
\begin{array}{cc} m^2(r)&dm(r)/dr\\dm(r)/dr&m^2(r)\end{array}
\right\} \;.
\end{eqnarray}
A very fast method for computing fluctuation determinants is based on a
theorem on functional determinants
\cite{Col} which can be generalized to
a coupled $(n \times n)$ system:

Let ${\bf f}(\nu,r)$ and ${\bf f^0}(\nu,r)$ denote the $(n \times n)$
matrices formed by $n$ linearly independent solutions
$f_i^\alpha(\nu,r)$ and $f_i^{\alpha 0}(\nu,r)$ of
\begin{equation}
\left( {\bf M_l} + \nu^2 \right)_{ij} f_j^\alpha(\nu,r) = 0
\end{equation}
and
\begin{equation}
\left( {\bf M_l^0} + \nu^2 \right)_{ij} f_j^{\alpha 0}(\nu,r) = 0
\end{equation}
respectively, with regular boundary conditions at $r=0$.
Here the latin lower index denotes the $n$ components
while the different solutions are labelled
by the Greek upper index. Let these solutions be normalized in
such a way  that
\begin{equation}
\lim_{r \to 0} {\bf f}(\nu,r) \left({\bf f^0}(\nu,r)
\right)^{-1}= {\bf 1}.
\end{equation}
Then the statement of the theorem is:
\begin{equation}
{\cal J}_l(\nu)\equiv
\frac{\det \left({\bf M_l} + \nu^2\right)}
{\det \left({\bf M_l^0} + \nu^2\right)} = \lim_{r \to \infty}
\frac{\det {\bf f}(\nu,r)}{\det {\bf f^0}(\nu,r)}.
\end{equation}
Here the determinants on the left-hand side are
determinants in functional
space while those on the right-hand side
are ordinary determinants of the
$n \times n$ matrices defined above. The theorem already
has been applied to the calculation of the one-loop
effective action of a single scalar field on a
bubble background in \cite{BaaKi,Baa2} and of a
fermion system at temperature $T=0$ on a similar background in
\cite{BaaSoSu}
which we refer to for more technical details.

In the numerical application the solutions $f_k^\alpha$ are written as
\cite{Baa}
\begin{equation}
f_k^\alpha(\nu,r) = \left( \delta_k^\alpha + h_k^\alpha(\nu,r) \right)
i_{l_k}(\nu r)
\end{equation}
with the boundary condition $h_k^\alpha(\nu,r) \to 0$ as $r \to 0$.
Of course the value $l_k$ depends  on the channel. This way one
generates a set of linearly independent solutions which near $r=0$
behave like the free
solution as required by the theorem which then takes the form
\begin{equation}
{\cal J}(\nu) = \lim_{r \to \infty} \det \left(\delta_k^\alpha+
h_k^\alpha(\nu, r) \right).
\end{equation}
The functions $h_k^\alpha(\nu,r)$ satisfy the differential equation
\begin{equation}
\frac{d^2}{dr^2} h_k^\alpha(\nu,r) + 2\left( \frac{1}{r} + \nu
\frac{i_{l_k}^\prime(\nu r)}{i_{l_k}(\nu r)}\right)
\frac{d}{dr} h_k^\alpha(\nu,r) = V_{kk^\prime}(r)
\Bigl( \delta_{k^\prime}^\alpha
+ h_{k^\prime}^\alpha(\nu,r)\Bigr) \frac{i_{l_{k^\prime}}(\nu r)}
{i_{l_k}(\nu r)}.
\end{equation}
This equation can easily be used
for generating the functions $h_k^\alpha$ order
by order in the potential $V$. Introducing the contribution of
order $k$ in the potential as ${\bf h}^{(k)}$ and and defining
${\bf h}^{\overline{(k)}}$ via
\[
{\bf h}^{\overline{(k)}} \equiv \sum_{j=k}^{\infty} {\bf h}^{(j)}
\]
as in \cite{BaaSoSu}, the relevant
 contribution $S\sub{eff}^{\overline{(3)}}$
is found to be
\begin{equation} \label{Matsum}
S\sub{eff}^{\overline{(3)}} = -\sum_{n=0}^{\infty}
{\cal K}^{\overline{(3)}}(\nu_n),
\end{equation}
where
\begin{eqnarray}
{\cal K}^{\overline{(3)}}(\nu)&=& \sum_{l=0}^{\infty} 6(2l+1)
\lim_{r \to \infty}
\left \{ \ln \det  \left( {\bf 1} +
{\bf h}^{\overline{(1)}}(\nu,r) \right)
\right. \nonumber\\
& & \left. - \tr \left( {\bf h}^{(1)}(\nu,r)
+ {\bf h}^{(2)}(\nu,r) - \frac{1}{2} \left[
{\bf h}^{(1)}(\nu,r)\right]^2 \right) \right\}.
\label{k3barnum}
\end{eqnarray}
The expression ${\cal K}^{\overline{(3)}}(\nu)$ was evaluated
by summing the partial waves computed numerically up to $l_{max}=30$
extrapolated for higher values of $l$ using an Ansatz
$a l^{-5} + b l^{-6} + c l^{-7}$. The asymptotic behaviour is supposed
to set in at values of $l >> \nu R$ where $R$ is the typical
radius of the bubble. Since $R$ has typical values of 20-40, this means
that for our $l_{max}$ the extrapolation becomes unreliable already
for $\nu$ of order $1$. We will discuss this point again below.
%--------------------------------------------------------------------
\subsection{Analytic approximation using the gradient expansion}
In \cite{BaaSoSu} an approximation of the gradient expansion type
has been given for the one-loop effective action at zero temperature
for the case of a massive fermion with Yukawa coupling to an external
scalar field. The
calculation to be done here is similar because
the Matsubara frequency $\nu_n$ enters in the same way as a mass term
and so the structure of the determinant is of the same type.

The logarithm of the determinant ${\cal J}(\nu)$ can be
written exactly as
\begin{equation}
\ln {\cal J}(\nu) = \Tr \ln \left( {\bf 1} + {\bf G_0} {\cal V}
\right)
\end{equation}
where the free Green function ${\bf G_0}$ is defined by
\begin{equation}
\left({\bf M} + \nu^2 \right) {\bf G_0} = {\bf 1}.
\end{equation}
Defining the Fourier transform of the potential ${\cal V}$ as
\begin{equation}
\widetilde {\cal V}({\bf q}) =
\int {\rm d}^3\!x\, {\cal V}({\bf x}) \exp (-i {\bf qx})
\end{equation}
the fluctuation determinant can be expanded as
\begin{equation}
\label{lnjnu}
\ln {\cal J}(\nu) = \sum_{k=1}^{\infty} \tr \frac{(-1)^{k+1}}{k}
\int \frac{{\rm d}^3\!p}{(2\pi)^3}\,
\prod_{j=1}^{k} \int \frac{{\rm d}^3\!q_j}{(2\pi)^3}\,
\frac{\widetilde{\cal V}({\bf q}_j)}{({\bf p} + {\bf Q}_j)^2+\nu^2}
(2\pi)^3 \delta^{(3)}({\bf Q}_k)
\end{equation}
where
\[
{\bf Q}_j = \sum_{l}^{j} {\bf q}_l.
\]
Expanding the denominators including terms up to order $\nu^{-2k-4}$
we get
\begin{eqnarray}
\prod_{j=1}^{k} \frac{1}{({\bf p} + {\bf Q}_j)^2+\nu^2} &\simeq&
\frac{1}{({\bf p}^2+\nu^2)^k} \left[ 1- \sum_{l=1}^{k}
\frac{{\bf Q}_l^2}{({\bf p}^2+\nu^2)} \right.\nonumber\\
& & \left. + \frac{4}{3} \frac{{\bf p}^2}{({\bf p}^2+\nu^2)^2} \left(
\sum_{l>l^\prime} {\bf Q}_l
{\bf Q}_{l^\prime} + \sum_{l=1}^{k}{\bf Q}_l^2
\right) \right].
\end{eqnarray}
As the potential is ${\cal V}=m_F^2\Phi^2+
 m_F\gamma\nabla \Phi$ we have
\begin{eqnarray}
\tr \prod_{j=1}^{k}
\widetilde{\cal V}({\bf q}_j) &\simeq& 4 m_F^{2k} \left[
\prod_{j=1}^{k} \widetilde{\Phi^2}({\bf q}_j) \right.\nonumber\\
& & \left. -
\frac{1}{m_F^2} \sum_{l>l^\prime} {\bf q}_l{\bf q}_{l^\prime}
\widetilde \Phi({\bf q}_l) \widetilde \Phi({\bf q}_{l^\prime})
\prod_{j\neq l,l^\prime} \widetilde{\Phi^2}({\bf q}_j) \right] \; .
\end{eqnarray}
After inserting these expansions in (\ref{lnjnu})
and transforming back to $x$-space the remaining integrations
(except of one space integration)
and the $k$-summation can be done. Of course we have
to omit those terms which are divergent. As
we are working in four dimensions these are
the terms with $k=1$ and $k=2$, they have been discussed in the
previous section.

Our final result written in the form of
${\cal K}(\nu) = 3 \ln {\cal J}(\nu)$ is
\begin{eqnarray}
\label{k3barana}
{\cal K}^{\overline{(3)}}(\nu) &\simeq&
\frac{3\nu^3}{2\pi} \intd{^3\!x} \left\{ -\frac{4}{3}\left[ \left(1+
\frac{m_F^2\Phi^2}{\nu^2}\right)^{\frac{3}{2}}
-1-\frac{3}{2}\frac{m_F^2\Phi^2}{\nu^2}-
\frac{3}{8}\frac{m_F^4\Phi^4}{\nu^4}\right] \right.\\
 &+& \left. \frac{m_F^2(\nabla\Phi)^2}{2\nu^4}\left[
\left(1-\left( 1+
\frac{m_F^2\Phi^2}{\nu^2}\right)^{-\frac{1}{2}}\right)
- \frac{m_F^2\Phi^2}{3\nu^2}\left( 1-\left( 1+
\frac{m_F^2\Phi^2}{\nu^2}\right)^{-\frac{3}{2}}\right )
\right]\right \}.\nonumber
\end{eqnarray}
This expression has to be evaluated using for $\Phi$ the numerical
bubble profiles. The accuracy is only limited by the accuracy of these
profiles and by that of the numerical integration. With our
numerical precision the results are reliable to at least
6 significant digits.

The approximate results for ${\cal K}^{\overline{(3)}}$ can be compared
with the exact numerical ones computed using
(\ref{k3barnum}). For the purpose of comparison we treat $\nu$ as a
continuous parameter. We display the exact and approximate results
in Figs. 1 and 2 for two typical bubble profiles, a small bubble
with $ y = 0.6$ and a large bubble corresponding
to $y = 0.3$ (see (\ref{ypsdef}) for the definition of
$y$). The analytic approximation is seen to describe the trend of the
exact results over the whole range.
The gradient expansion is expected to converge at large
$\nu$. This expectation is substantiated by the exact
numerical results in the region where they are reliable.
It is seen, however, that the exact results start dropping
 off at values of $\nu \simeq 1 - 2$; as mentioned above
this is related to the fact that the
convergence of the partial wave
summation becomes poorer with increasing $\nu$. Since
the values of $\nu$ relevant
for the Matsubara frequency summation (\ref{Matsum})
are $\nu \ge \pi T \approx 9$
( in our units $g\tilde v$) we have to rely on the gradient
expansion in computing the finite temperature effective action.

%--------------------------------------------------------------------

\section{Results}
\setcounter{equation}{0}

We have computed the finite temperature fermionic effective action
for Higgs masses of $60, 70 $ and $ 80$ GeV and for top quark
masses $m_t = 160, 170$ and $180$ GeV. These results are
given in Tables 1 to 3. As mentioned in
section 2 we have considered only
the contribution of the top quark since lighter quarks and leptons
will give negligible contributions. Their contribution has already
been dismissed in the basic high temperature action (\ref{htac}) and
including them would be inconsistent.

For each set of mass parameters we have determined the bubble
profiles for various values of the variable $y$ defined in
(\ref{ypsdef}); $y$ determines the temperature and the bubble
action. We give separately the renormalized first and second
order contributions $\Delta S\sub{ren}^{(1+2)}$, Eq. (\ref{DeltaS12})
 determined in section $3$
and the finite sum of all higher order contributions
$S\sub{eff}^{\overline{(3)}}$ whose computation was discussed in the
previous section. It is given by the analytic expression
Eq. (\ref{k3barana}), inserted into the Matsubara sum (\ref{Matsum}).
The total one-loop effective
action, reduced by the terms included already
in the `classical' effective potential is given, of course,
by the sum $\Delta S\sub{eff}^{(1+2)} + S\sub{eff}^{\overline{(3)}}$.

The fermion determinant is seen to yield a negative contribution
to the effective action, which means that bubble nucleation is
enhanced by this contribution (cf. Eqs. (\ref{rate}) and
(\ref{sefff})). It is interesting to analyze the
relative importance of the various contributions. The $\overline{(3)}$
part given in Eq. (\ref{k3barana})
is relatively small, also the zero-derivative part of the
leading order $(1+2)$ contributions in Eq.
(\ref{DeltaS12}).
This is not unexpected, as the zero-derivative part was already
included into the effective potential, so its leading first and
second order contributions are subtracted.
The dominant contribution is the gradient term in Eq.
(\ref{DeltaS12}), the finite part of the wave function renormalization.
It is also displayed in Tables 1 to 3 in the colums labelled
$\Delta^{(1+2)}_{\rm grad}$.
This means that the top quark
contribution can be described, essentially, by local terms in analytic
form: those already contained
in the effective potential and the finite part of
the wave function renormalization; this is very convenient because
these terms can be incorporated into the basic action from which the
bubble profile is computed. If this is done, the remaining corrections
can be expected to be very small. In a selfconsistent determination
of the bubble profiles such an action can be expected to yield an
excellent first approximation.
\newpage

%******************************************************** References

\newpage
\section*{Table Captions}
{\bf Table 1} ~Corrections to the effective action at finite
temperature for $m_H=60$ GeV. $\epsilon$ and $y$ are defined
in Eq. (\ref{epsdef})
and (\ref{ypsdef}); $\Delta S\sub{ren}^{(1+2)}$ is
defined by Eqs. (\ref{DeltaS12}), $\Delta S\sub{grad}^{(1+2)}$
is the gradient part as explained in section 6; $S^{\overline{(3)}}$
is given by Eqs. (\ref{Matsum}) and (\ref{k3barana});
$\tilde S$, the classical action of Eq. (\ref{Stdef}) is given
for comparison. \\ \\
{\bf Table 2} ~Corrections to the effective action at finite
temperature for $m_H=70$ GeV. Definitions as in Table 1. \\ \\
{\bf Table 3} ~Corrections to the effective action at finite
temperature for $m_H=80$ GeV. Definitions as in Table 1.\\ \\
\section*{Figure Captions}
{\bf Fig. 1} ~The loop expansion of the effective action. The lines
represent the propagators and the dots indicate the vertices
V(x).\\ \\
{\bf Fig. 2} ~Results of the numerical computation compared to the
analytic approximation for $ y = 0.3$,~$m_H = 60 $ GeV and
$m_t = 170$ GeV. The dots interpolated by a dotted line represent
the numerical results, the solid line is the analytic approximation
of Eq. (\ref{k3barnum})\\ \\
{\bf Fig. 3} ~The same as Figure 2 for $y = 0.6$.
\newpage
\begin{table}
\begin{center}
\begin{tabular}{|c|c|c|c|c|c|c|c|}
\hline
& & & & & & & \\[-0.4cm]
$m_t [\mbox{GeV}]$ & T $[$GeV$]$ & $\epsilon$ & y &
$\triangle S\sub{ren}^{(1+2)}$ & $\triangle S\sub{grad}^{(1+2)} $ &
$S\sub{eff}^{\overline{(3)}}$ & $\tilde S$\\
\hline
160 & 94.894 & 1.867 & 0.2 & -23.160 & -23.250 & -18.923 & 308.02\\
 & 94.854 & 1.800 & 0.3 & -9.637 & -9.753 & -3.774 & 132.30\\
 & 94.682 & 1.600 & 0.6 & -1.685 & -1.693 & -0.0781 & 24.847\\
\hline
170 & 94.5288 & 1.867 & 0.2 & -23.752 & -23.695 & -20.418 & 278.47\\
 & 94.495 & 1.800 & 0.3 & -9.973 & -10.091 & -4.184 & 121.40\\
 & 94.347 & 1.600 & 0.6 & -1.734 & -1.742 & -0.0854 & 22.686\\
\hline
180 & 94.6605 & 1.867 & 0.2 & -24.558 & -24.291 & -20.929 & 251.60\\
 & 94.631 & 1.800 & 0.3 & -10.002 & - 10.114 & -4.110 & 107.27\\
 & 94.5056 & 1.600 & 0.6 & -1.757 & -1.766 & -0.0855 & 20.245\\
\hline
\end{tabular}

\vspace{8mm}

\Large Table 1

\end{center}
\end{table}

\begin{table}
\begin{center}
\begin{tabular}{|c|c|c|c|c|c|c|c|}
\hline
& & & & & & & \\[-0.4cm]
$m_t [\mbox{GeV}]$ & T $[$GeV$]$ & $\epsilon$ & y &
$\triangle S\sub{ren}^{(1+2)}$ & $\triangle S\sub{grad}^{(1+2)}$ &
$S\sub{eff}^{\overline{(3)}}$ & $\tilde S$\\
\hline
160 & 106.015 & 1.867 & 0.2 & -18.933 & - 18.740 & -6.746 & 217.14\\
 & 105.980 & 1.800 & 0.3 & -7.956 & - 8.006 & -1.384 & 94.690\\
 & 105.826 & 1.600 & 0.6 & -1.387 & -1.391 & -0.0282 & 17.630\\
\hline
170 & 104.7850 & 1.867 & 0.2 & -20.091 & -19.899 & -8.412 & 207.39\\
 & 104.7545 & 1.800 & 0.3 & -8.438 & -8.500 & -1.726 & 90.411\\
 & 104.6202 & 1.600 & 0.6 & -1.470 & -1.475 & -0.0352 & 16.830\\
\hline
180 & 104.094 & 1.867 & 0.2 & -20.897 & -20.743 & -9.602 & 193.72\\
 & 104.0674 & 1.800 & 0.3 & -8.751 & -8.799 & -1.946 & 83.921\\
 & 103.951 & 1.600 & 0.6 & -1.522 & -1.528 & -0.0396 & 15.632\\
\hline
\end{tabular}

\vspace{8mm}

{\Large Table 2}
\end{center}
\end{table}

\begin{table}
\begin{center}
\begin{tabular}{|c|c|c|c|c|c|c|c|}
\hline
& & & & & & & \\[-0.4cm]
$m_t [\mbox{GeV}]$ & T $[$GeV$]$ & $\epsilon$ & y &
$\triangle S\sub{ren}^{(1+2)}$ & $\triangle S\sub{grad}^{(1+2)}$ &
$S\sub{eff}^{\overline{(3)}}$ & $\tilde S$\\
\hline
160 & 117.541 & 1.867 & 0.2 & -15.433 & -15.177 & -2.512 & 156.45\\
 & 117.510 & 1.800 & 0.3 & -6.483 & -6.483 & -0.514 & 68.050\\
 & 117.374 & 1.600 & 0.6 & -1.136 & -1.139 & -0.0107 & 12.708\\
\hline
170 & 115.4848 & 1.867 & 0.2 & -16.747 & -16.411 & -3.388 & 153.28\\
 & 115.4573 & 1.800 & 0.3 & -6.985 & -6.984 & -0.688 & 66.460\\
 & 115.337 & 1.600 & 0.6 & -1.222 & -1.225 & -0.0142 & 12.410\\
\hline
180 & 114.0042 & 1.867 & 0.2 & -17.696 & -17.287 & -4.163 & 146.18\\
 & 113.9801 & 1.800 & 0.3 & -7.387 & -7.386 & -0.852 & 63.649\\
 & 113.874 & 1.600 & 0.6 & -1.287 & -1.290 & -0.0174 & 11.846\\
\hline
\end{tabular}

\vspace{8mm}

{\Large Table 3}
\end{center}

\end{table}

\clearpage
\begin{figure}
\begin{center}
\setlength{\unitlength}{1.2cm}
\begin{picture}(12,3)
\thicklines
\put(0.3,1.3){\Large $S_{\rm eff}= \quad -\frac{1}{2}$}
\multiput(3.5,1.5)(3,0){3}{\circle{1.2}}
\multiput(3.5,0.9)(3,0){3}{\circle*{0.2}}
\put(6.5,2.1){\circle*{0.2}}
\put(8.98,1.8){\circle*{0.2}}
\put(10.02,1.8){\circle*{0.2}}
\put(4.6,1.3){\Large $+\quad \frac{1}{4}$}
\put(7.6,1.3){\Large $- \quad \frac{1}{6}$}
\put(10.6,1.3){\Large $+ \dots$}
\end{picture}

\vspace{8mm}

 \Large Figure 1
\end{center}
\end{figure}

\begin{figure}
\centerline{
\mbox{\epsfxsize=15cm
\epsfbox{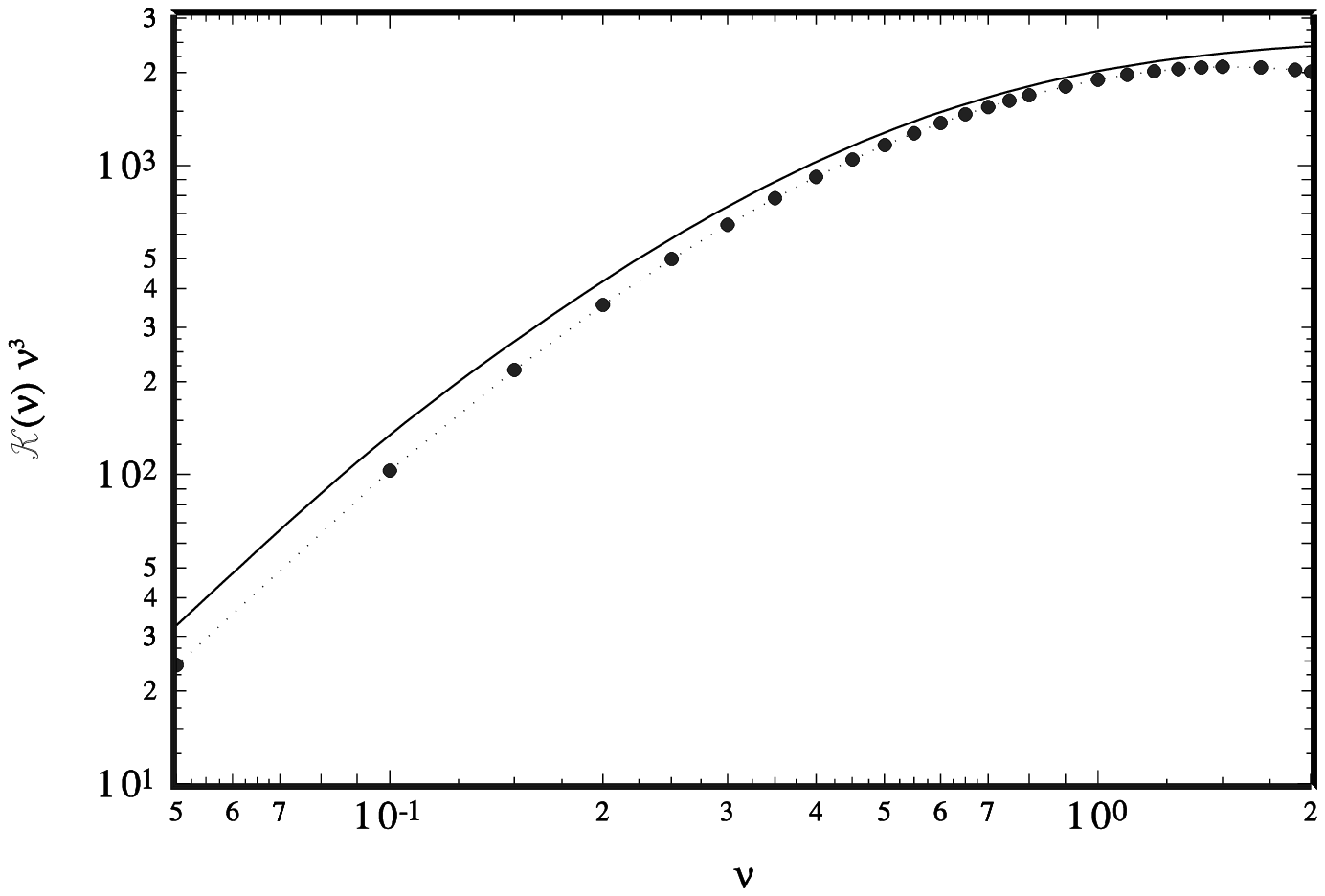}}
}

\vspace{8mm}

\centerline{\Large Figure 2}

\end{figure}
\begin{figure}
\centerline{
\mbox{\epsfxsize=15cm
\epsfbox{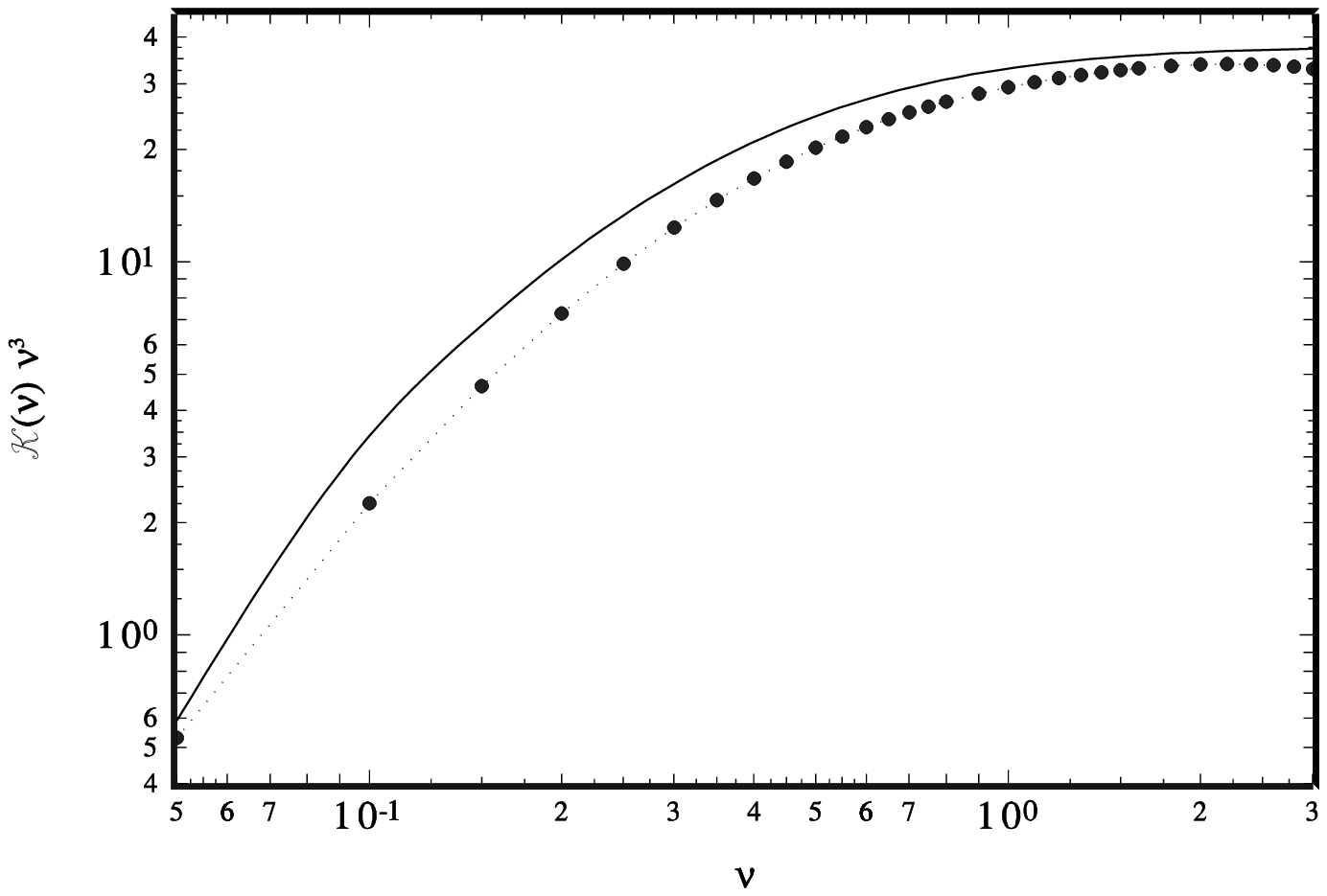}}
}

\vspace{8mm}

\centerline{\Large Figure 3}
\end{figure}

\begin{thebibliography}{}

\bibitem{Sin} see `Electroweak physics and the Early Universe'
Eds. F. Freire and J. Rom\~ao, \\ Plenum Publ. Corp.,
N. Y., 1994,
for a recent account of the state of art.

\bibitem{KuRuSha} V. Kuzmin, V. Rubakov and M. Shaposhnikov,
Phys. Lett. {\bf B155}, 36 (1985).

\bibitem{CoKaNe} for a review see A. Cohen, D. Kaplan and
A. Nelson, Ann. Rev. Nucl. Part. Sci.
{\bf 43}, 27 (1993).

\bibitem{FaSha}G. Farrar and M. Shaposhnikov, Phys. Rev. Lett.
{\bf 70}, 2833 (1993), {\bf 71}, 210(E) (1993), CERN preprints
CERN-TH 6732-93 and CERN-TH 6734-93.

\bibitem{GHOP}
M. B. Gavela, M. Lozano, J. Orloff and O. Pene, Nucl. Phys.
{\bf B430}, 345 (1994); M. B. Gavela, P. Hernandez, J. Orloff,
O. Pene and C. Quimbay, Nucl. Phys. {\bf B430}, 382 (1994).

\bibitem{BFHW} W. Buchm\"uller, Z. Fodor, T. Helbig and
D. Walliser, Ann. Phys. {\bf 234}, 260 (1994).

\bibitem{FKRSh1} K. Farakos, K. Kajantie, K. Rummukainen and
M. Shaposhnikov, Nucl. Phys. {\bf B425}, 67 (1994).

\bibitem{KRSh} K. Kajantie, K. Rummukainen and M. Shaposhnikov,
Nucl. Phys. {\bf B407}, 356 (1993).

\bibitem{FHJJMC} Z. Fodor, J. Hein, K. Jansen, A. Jaster,
I. Montvay and F. Czikor, Phys. Lett. {\bf B334}, 405 (1994).

\bibitem{FKRSh2} K. Farakos, K. Kajantie, K. Rummukainen and
M. Shaposhnikov, Phys. Lett. {\bf B336}, 494 (1994).

\bibitem{TuWeWi} M. S. Turner, E. J. Weinberg and
L. M. Widrow, Phys. Rev. {\bf D46}, 2384 (1992).

\bibitem{LiLeTu} B. H. Liu, L. McLerran and N. Turok,
Phys. Rev. {\bf D46}, 2688 (1992).

\bibitem{DLHLL} M. Dine, R. G. Leigh, P. Huet, A. Linde
and D. Linde, Phys. Rev. {\bf D46}, 550 (1992).

\bibitem{La} J. S. Langer,  Ann. Phys. (N. Y.)
{\bf 41}, 108 (1967); {\it ibid.} {\bf 54}, 258 (1969).

\bibitem{Co}S. Coleman, Phys. Rev. {\bf D15}, 2929 (1977).

\bibitem{CaCo}G. Callan and S. Coleman,
Phys. Rev. {\bf D16}, 1762 (1977).

\bibitem{KriLaSch} J. Kripfganz, A. Laser and M. G. Schmidt,
Nucl. Phys. {\bf B433}, 467 (1995).

\bibitem{Baa2} J.~Baacke, Preprint DO-TH-95/04, March 1995.

\bibitem{DyaGoe} D. Dyakonov, M. Polyakov, P. Sieber, J. Schaldach
and K. Goeke,
Phys. Rev. {\bf D49}, 6964 (1994).

\bibitem{CoWe} S. Coleman and E. Weinberg,
Phys. Rev. {\bf D7}, 1888 (1973).

\bibitem{KiLi} D. A. Kirshnits and A. D. Linde,
Ann. Phys. (N.Y.) {\bf101}, 195 (1976).

\bibitem{Sha} M. E. Shaposhnikov, Nucl. Phys.
{\bf B287}, 757 (1987); {\it ibid.} {bf B 299}, 581 (1987).

\bibitem{AnHa} G. W. Anderson and L. J. Hall,
Phys. Rev. {\bf D45}, 2685 (1991).

\bibitem{Wei} E. J. Weinberg, Phys. Rev. {\bf D47}, 4614 (1993).

\bibitem{Af}  I. Affleck, Phys. Rev. Lett. {\bf 46},
388 (1981).

\bibitem{Li} A. D. Linde,  Nucl. Phys. {\bf B216},421 (1983).

\bibitem{BaaKi} J. Baacke and V. G. Kiselev, Phys. Rev.
{\bf D48}, 5648 (1993).

\bibitem{DoJa} L. Dolan and R. Jackiw,
Phys. Rev. {\bf D9}, 3320 (1974).

\bibitem{Col} see e.~g. S.~Coleman, {\em The Uses of Instantons\/}, in
{\em The Aspects of Symmetry\/}, Cambridge University Press 1985.

\bibitem{BjDr} J. D. Bjorken and S. Drell,
{\it Relativistic Quantum Mechanics}, McGraw-Hill Book
Company, New York 1964.

\bibitem{BaaSoSu} J. Baacke, H. So and A. Suerig,
Z. Phys. {\bf C63}, 689 (1994).

\bibitem{Baa} J. Baacke, Z. Phys. {\bf C53}, 407 (1992).



\end{thebibliography}
\end{document}